\documentclass[aps,twocolumn,showpacs,preprintnumbers,amsmath,amssymb]{revtex4}

\usepackage{graphicx}
\usepackage[dvips]{color}
\usepackage{mathrsfs}

\begin{document}


\title{Critical velocity of flowing supersolids of dipolar Bose gases in optical lattices}
\author{Ippei Danshita$^{1}$}
\altaffiliation{Present address: RIKEN, Wako, Saitama 351-0198, Japan}
\author{Daisuke Yamamoto$^{2}$}
\affiliation{
{$^1$Department of Physics, Faculty of Science, Tokyo University of Science,
Shinjuku-ku, Tokyo 162-8601, Japan}
\\
{$^2$Department of Physics, Waseda University, Shinjuku-ku, Tokyo 169-8555, Japan}
}

\date{\today}

\begin{abstract}
We study superfluidity of supersolid phases of dipolar Bose gases in two-dimensional 
optical lattices.
We perform linear stability analyses for the corresponding dipolar Bose-Hubbard 
model in the hardcore boson limit to show that a supersolid can have stable superflow 
until the flow velocity reaches a certain critical value.
The critical velocity for the supersolid is found to be significantly smaller than that for 
a conventional superfluid phase.
We propose that the critical velocity can be used as a signature to identify the superfluidity
of the supersolid phase in experiment.

\end{abstract}


\maketitle
\section{introduction}
New possibilities to explore exotic quantum phases have been pioneered by recent 
experimental advances in creating dipolar ultracold gases~\cite{lahaye-09,baranov-08}, 
such as the realization of a condensate of $^{52}$Cr atoms with strong magnetic dipole 
moments~\cite{griesmaier-05,lahaye-07} and heteronuclear polar 
molecules~\cite{ni-08,ospelkaus-09}.  
Thanks to the long-range nature and anisotropy of the dipole-dipole interactions, 
various quantum phases have been predicted to emerge, including fermionic superfluids 
(SF) with $p$-wave pairing~\cite{baranov-08,bruun-08}, 
Haldane-Bose insulators~\cite{torre-06}, and 
supersolids (SS)~\cite{goral-02,kovrizhin-05,yi-07,danshita-09,sansone-10,pollet-10}.

Of particular interest are SS phases, in which both diagonal (crystalline) 
and off-diagonal (superfluid) long-range orders coexist~\cite{andreev-69}. 
Although non-classical rotational inertia, one of the signatures of superfluidity, 
was experimentally observed in solid helium~\cite{kim-04}, it has been more 
reasonably interpreted by other mechanisms, such as superfluidity of grain boundaries 
and dislocations~\cite{sasaki-06,pollet-07}. 
On the other hand, quantum Monte Carlo (QMC) simulations have shown the presence 
of SS phases in Bose-Hubbard systems with long-range 
interactions~\cite{sansone-10,pollet-10,qmc-series,batrouni-00}. 
Since QMC analyses of the Bose-Hubbard model quantitatively agree with experiments 
of ultracold gases in optical lattices, the SS phases are expected to be found in 
the context of dipolar Bose gases loaded into optical lattices.

In order to verify the existence of SS in experiments of ultracold gases, one has to clarify 
observables to identify the superfluidity and the crystalline order of the SS phases. 
It is well-known that the crystalline order can be identified by the static 
structure factor, which has been observed in cold atom experiments 
using the Bragg scattering techniques~\cite{steinhauer-02,ernst-09}.
On the other hand, a sharp interference peak in the time-of-flight image following the expansion
of a gas is often used as an indirect indication of the superfluidity of Bose gases in optical 
lattices~\cite{greiner-02}.
However, the sharp interference peak identifies the presence of a Bose-Einstein condensate,
but does not necessarily mean the superfluidity.
For instance, although a non-interacting Bose gas forms a condensate at sufficiently low
temperature, it is not a SF in the sense that its critical velocity is zero~\cite{pitaevskii}.
Moreover, while most previous theoretical work calculated the SF fraction as 
a characteristic of the superfluidity~\cite{sansone-10,pollet-10,qmc-series,batrouni-00}, 
so far no experiment has succeeded in measuring the SF fraction in cold atom 
systems~\cite{footnote}, in contrast to helium systems where the SF fraction, corresponding to 
the non-classical rotational inertia, can be easily measured with a torsional 
oscillator~\cite{kim-04,sasaki-06}.
Instead, the superfluidity of weakly-~\cite{sarlo-05} and strongly-interacting Bose 
gases~\cite{mun-07}, and fermionic SF across the BEC-BCS 
crossover~\cite{miller-07} has been demonstrated in a moving optical lattice 
by measuring the critical velocity above which superflow breaks down.

In this paper, we propose that the superfluidity of SS phases can also be experimentally 
identified using a moving optical lattice. 
Performing linear stability analyses for polarized dipolar hardcore bosons
in a two-dimensional (2D) moving optical lattice, we show that superflow of SS states is stable 
until the flow momentum exceeds a certain finite value. 
It is found that the critical momenta for the dynamical instability in the SS phases 
are remarkably smaller than that for a standard SF with no density wave order.
We argue that the critical momenta can be experimentally measured with currently 
available techniques.

The remainder of the paper is organized as follows. 
In Sec.~\ref{sec:model}, we introduce our model Hamiltonian describing hardcore bosons with 
dipole-dipole interaction in a 2D optical lattice.
In Sec.~\ref{sec:MF}, we explain our formulation of the problem based on a mean-field theory.
In Sec.~\ref{sec:GSPD}, we calculate the ground-state phase diagram.
In Sec.~\ref{sec:Vc}, we perform linear stability analyses to obtain the critical velocities for 
Landau and dynamical instabilities in the SF and SS phases.
In Sec.~\ref{sec:Sum}, we summarize our results and briefly discuss two recent experiments 
exploring SS phases in cold atom systems~\cite{vengalattore-10,baumann-10}.

\section{Model}
\label{sec:model}
We consider a system of $N$ bosons interacting with onsite and dipole-dipole
interactions in a 2D optical lattice.
The dipoles are assumed to be polarized to the direction perpendicular to the lattice plane.
This system can be well-described by the dipolar Bose-Hubbard model~\cite{goral-02},
\begin{eqnarray}
\hat{H} &=& 
-J\sum_{\langle j,l \rangle}
(\hat{a}^{\dagger}_{j} \hat{a}_{l}+{\rm h.c.})
+ \frac{U}{2} \sum_{j}
\hat{n}_{j} (\hat{n}_{j} -1)
\nonumber \\
&&
+\sum_{ j < l }
V_{j l} \hat{n}_{j}\hat{n}_{l}
-\mu \sum_j \hat{n}_j,
\label{eq:hamiltonian}
\end{eqnarray}
where $\hat{a}^{\dagger}_{j}$ is the boson creation operator at site $j$,
$\hat{n}_j=\hat{a}^{\dagger}_{j} \hat{a}_{j}$, $J$ is the hopping, and 
$U$ is the onsite interaction.
The chemical potential $\mu$ controls the filling factor $n\equiv N/M$, 
where $M$ is the total number of lattice sites.
$\langle j,l \rangle$ represents nearest-neighbor pairs of lattice sites.
The long-range part of the dipole-dipole interaction is well-approximated as 
$V_{jl}=V d^3|{\bf r}_j-{\bf r}_l|^{-3}$, where $j_x$ and $j_y$ are integers and 
$d$ is the lattice spacing.
In experiments, the ratios $J/V$ and $U/V$ can be controlled by varying the lattice depth
and using the Feshbach resonance~\cite{lahaye-07}.

In the hardcore boson limit ($U\rightarrow \infty$), Eq.~(\ref{eq:hamiltonian}) can be
mapped onto the following spin-1/2 Hamiltonian,
\begin{eqnarray}
\hat{H}_{\rm s}=
-J\sum_{\langle j,l \rangle}
(\hat{S}^{+}_{j} \hat{S}^{-}_{l}+{\rm h.c.})
+ \sum_{ j < l } V_{jl} \hat{S}^{z}_{j}\hat{S}^{z}_{l}
-h \sum_j \hat{S}^{z}_j,
\label{eq:spinH}
\end{eqnarray}
where the spin operators are related to the operators of the hardcore boson as
$\hat{S}_{j}^{-}=\hat{a}_j$  and $\hat{S}_j^z = \hat{n}_j -1/2$.
%
$h=\mu-2C_{+} V$
%
is the shifted chemical potential, where
\begin{eqnarray}
C_{\pm} &=& \sum_{\alpha=1}^{\infty}
\frac{(\pm 1)^{\alpha - 1}}{\alpha^{3}}
+
\sum_{\alpha_x,\alpha_y=1}^{\infty}
\frac{ (\pm 1)^{\alpha_x + \alpha_y -1} }
{(\alpha_x^2+\alpha_y^2)^{3/2}}.
\end{eqnarray}
These coefficients include the long-range nature of the dipole-dipole interaction and 
their numerical values are $C_{+}\simeq 2.258$ and $C_{-}\simeq 0.6615$.
The dynamics of the spin model Eq.~(\ref{eq:spinH}) is described by the Heisenberg
equation of motion for $\hat{S}_j^{+}$ (we set $\hbar = 1$),
\begin{eqnarray}
i\frac{d}{dt}\hat{S}_j^{+} = -2J \hat{S}_j^z \sum_{\langle l \rangle} \hat{S}_l^{+}
- \hat{S}_j^{+}\sum_{l\neq j} V_{jl} \hat{S}_l^{z} + h \hat{S}_j^{z},
\label{eq:Heq}
\end{eqnarray}
where $\langle l \rangle$ represents the nearest-neighboring sites to site $j$.

\section{Mean-field theory}
\label{sec:MF}
Our formulation of the problem is based on a mean-field theory, in which
a many-body wave function is approximated as a product of 
the local spin coherent states,
\begin{eqnarray}
|\Phi_{\rm MF}\rangle = 
\prod_j \left( e^{-\frac{i\varphi_j}{2}}\cos\frac{\theta_j}{2} |\!\uparrow\rangle_j
+ e^{\frac{i\varphi_j}{2}}\sin\frac{\theta_j}{2} |\!\downarrow\rangle_j 
\right),
\end{eqnarray}
where $\theta_j$ and $\varphi_j$ are the elevation and azimuthal angles of the spin 
direction at site $j$.
Replacing $\hat{S}_j^{z}$ and $\hat{S}_j^{+}$ with 
$\langle \hat{S}_j^{z} \rangle = \frac{1}{2}\cos \theta_j$ 
and $\langle \hat{S}_j^{+} \rangle = \frac{1}{2}e^{i \varphi_j} \sin \theta_j$ in 
Eqs.~(\ref{eq:spinH}) and (\ref{eq:Heq}),
we obtain the mean-field energy $\mathcal{H}_0 \equiv \langle \hat{H}_{\rm s} \rangle$
given by
\begin{eqnarray}
\mathcal{H}_0 &=& 
- \frac{J}{2} \sum_{\langle j,l \rangle} \sin \theta_j \sin \theta_l \cos\varphi_{jl}
+\frac{1}{4} \sum_{j < l} V_{jl} \cos \theta_j \cos \theta_l 
\nonumber \\
&&- \frac{h}{2}\sum_j \cos \theta_j
\label{eq:MFenergy}
\end{eqnarray}
and the classical equations of motion for $\theta_j$ and $\varphi_j$:
\begin{eqnarray}
\frac{d \theta_j}{dt} = 
J \sum_{\langle l \rangle} \sin \theta_l \sin \varphi_{jl},
\label{eq:classical1}
\end{eqnarray}
\begin{eqnarray}
\frac{d \varphi_j}{dt} = 
J \sum_{\langle l \rangle} \frac{\sin \theta_l \cos \theta_j}{\sin \theta_j} \cos \varphi_{jl}
+ \frac{1}{2} \sum_{l\neq j} V_{jl} \cos \theta_l - h,
\label{eq:classical2}
\end{eqnarray}
where $\varphi_{jl}\equiv \varphi_j - \varphi_l$.
Equation (\ref{eq:classical1}) is the continuity equation while Eq.~(\ref{eq:classical2}) 
corresponds to the Josephson's acceleration equation.
The hardcore boson density $n_j \equiv \langle \hat{n}_j \rangle$, 
the density $n_j^{\rm con} = | \langle \hat{a}_j \rangle |^2$,
and the phase $\phi_j = \arg(\langle \hat{a}_j \rangle)$
of the condensate wave function, and the spin angles are related by 
$n_j  = (\cos \theta_j +1)/2$, $n_j^{\rm con}=n_j(1-n_j)$, and $\phi_j = -\varphi_j$.

Considering small-amplitude oscillations around a steady solution,
we write the solution of Eqs.~(\ref{eq:classical1}) and (\ref{eq:classical2}) in the form
\begin{eqnarray}
\theta_j(t) = \bar{\theta}_j + \delta \theta_j e^{-i \omega t},
\,\,
\varphi_j(t) = \bar{\varphi}_j + \delta \varphi_j e^{-i \omega t},
\label{eq:fluct}
\end{eqnarray}
where $\omega$ is the frequency of the normal mode.
Substituting Eq.~(\ref{eq:fluct}) into Eqs.~(\ref{eq:classical1}) and (\ref{eq:classical2})
and neglecting the terms higher than the first order with respect to $\delta \theta_j$ 
and $\delta \varphi_j$,
we obtain the equations for a steady state
\begin{eqnarray}
&&\sum_{\langle l \rangle} \sin \bar{\theta}_l \sin \bar{\varphi}_{jl}=0,
\label{eq:static1}
\\
%
%
h &=& J \sum_{\langle l \rangle} 
       \frac{\sin \bar{\theta}_l \cos\bar{\theta}_j}{\sin \bar{\theta}_j} \cos \bar{\varphi}_{jl}
      +\frac{1}{2}\sum_{l\neq j} V_{jl} \cos \bar{\theta}_l,
\label{eq:static2}
\end{eqnarray}
and the linearized equations for fluctuations
\begin{eqnarray}
- i \omega \delta \theta_j 
\!\!\!&=&\!\!\!
J\! \sum_{\langle l \rangle} \left[ \delta \varphi_{jl} \sin \bar{\theta}_l \cos \bar{\varphi}_{jl}
\!+\! \delta \theta_l \cos \bar{\theta}_l \sin \bar{\varphi}_{jl} 
\right], 
\label{eq:bogo1} \\
%
%
- i  \omega \delta \varphi_j 
\!\!\!&=&\!\!\!
J \sum_{\langle l \rangle} \left[
\left(
\delta \theta_l \frac{\cos \bar{\theta}_j \cos \bar{\theta}_l}{\sin \bar{\theta}_j}
- \delta \theta_j \frac{\sin \bar{\theta}_l}{\sin^2 \bar{\theta}_j}
\right)\cos \varphi_{jl}
\right.
\nonumber \\
&&
\!\!\!\!\!\!\!\!\!\!\!\!\!\! \left.
- \delta \varphi_{jl}
\frac{\cos \bar{\theta}_j \sin \bar{\theta}_l \sin \bar{\varphi}_{jl}}{\sin \bar{\theta}_j} 
\right]
 - \frac{1}{2}\sum_{l \neq j} V_{jl} \delta \theta_l \sin \bar{\theta}_l.
\label{eq:bogo2}
\end{eqnarray}
The excitation energy $\omega$ calculated from Eqs.~(\ref{eq:bogo1}) and (\ref{eq:bogo2}) 
coincides with that calculated by linear spin-wave theory~\cite{scalettar-95}.
Stability of a steady solution can be discriminated by $\omega$.
The appearance of excitations with $\omega < 0$ signals the Landau instability (LI),
while the appearance of excitations with ${\rm Im}[\omega] \neq 0$ signals the 
dynamical instability (DI), which means exponential growth of the fluctuations in time.
The linear stability analyses on the basis of Eqs.~(\ref{eq:bogo1}) and (\ref{eq:bogo2}) allow 
us to calculate the critical velocity of superflow.

Previous theoretical analyses have shown that the mean-field theory fails to describe even
qualitatively the ground-state phase diagram of the hardcore Bose-Hubbard model 
with nearest- and next-nearest-neighbor interactions due to strong quantum 
fluctuations~\cite{scalettar-95,batrouni-00}.
More specifically, while the mean-field theory predicts the presence of a stable checkerboard
supersolid (CSS) phase~\cite{scalettar-95}, this SS phase is unstable towards phase separation
according to accurate QMC simulations.
However, the mean-field theory is qualitatively valid for our dipolar hardcore Bose-Hubbard
model of Eq.~(\ref{eq:spinH}), because the long-range nature of the dipolar interaction 
significantly suppresses quantum fluctuations~\cite{yamamoto-10}.
We will indeed show in the next section that the mean-field phase diagram for 
Eq.~(\ref{eq:spinH}) qualitatively agrees with the recent QMC results of 
Ref.~\cite{sansone-10}.
We do not argue that the mean-field theory can provide {\it quantitatively} correct results,
but that it is useful for gaining qualitative features and analytical insights of the critical velocity.
Notice that so far the QMC methods have not succeeded in calculating the critical velocity.

\section{Phase diagram for $K=0$}
\label{sec:GSPD}
\begin{figure}[t]
\includegraphics[scale=0.45]{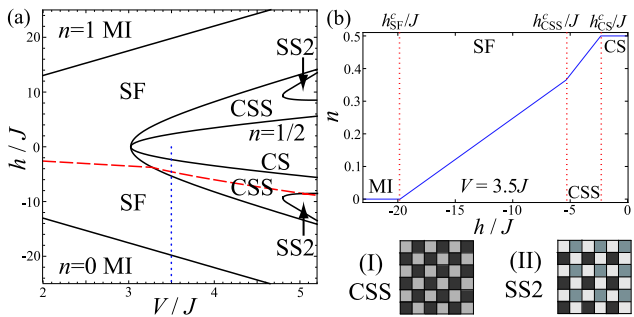}
\caption{\label{fig:PD}
(color online) (a) Ground-state phase diagram of the hardcore Bose-Hubbard model
with the dipole-dipole interaction in the $(V/J,h/J)$-plane, where MI, SF, CSS, CS, and 
SS2 phases are present.
The dashed red line represents the contour of $n=0.4$.
(b) The filling factor $n$ as a function of $h/J$.
The dotted red lines locate the boundaries between the different phases.
The dipolar interaction is fixed to be $V=3.5J$, which is indicated by the dotted blue line in (a).
(I) and (II) Schematic pictures of the CSS and SS2 phases.
}
\end{figure}
Within the mean-field theory, let us first calculate the ground-state phase diagram of 
Eq.~(\ref{eq:spinH}) in the case that the system does not have a supercurrent, i.e. 
$\bar{\varphi}_j$ is constant.
Without loss of generality, we can consider the spins ordered in the $XZ$ plane 
($\bar{\varphi}_j = 0$).
To obtain analytical expressions of phase boundaries, we assume 
that the stationary solution satisfies $\bar{\theta}_j = \theta_{A(B)}$ for 
$j_x + j_y \in {\rm even (odd)}$~\cite{matsuda-70,scalettar-95}.
Under this two-sublattice ansatz, we can describe checkerboard solid (CS), 
SF, CSS, and  Mott insulator (MI) phases.
The CS phase, termed as the N\'eel state in the language of the spin model, is 
an incompressible insulating phase at half filling with a density wave order
whose ordering vector is ${\bf k}_{\pi}=(\pi/d,\pi/d)$.
This phase is favored in the region of $J, |h| \ll V$, where the antiferromagnetic
Ising term is dominant in Eq.~(\ref{eq:spinH}).
The SF phase is characterized by uniform density and finite condensate fraction, 
i.e. $n_j = n$ and $n_j^{\rm con} \neq 0$.
The latter condition reflects the existence of the off-diagonal long-range order.
This phase corresponds to a canted ferromagnetic state in the spin system and 
is favored when $J/V$ is large or $|h|/V$ is moderately large.
The CSS phase possesses both the SF and checkerboard density-wave orders.
Recent quantum Monte-Carlo simulations of Eq.~(\ref{eq:spinH}) in Ref.~\cite{sansone-10}
have shown that the CSS phase is indeed present in the intermediate region between the SF and 
CS phases.
MI is an incompressible phase with $n=0$ or $1$, which corresponds in the spin language 
to a fully polarized magnetic phase in a strong magnetic field $|h|$.
In terms of the spin angle, the conditions for the different phases to emerge is as follows:
\begin{eqnarray}
&& \cos \theta_A = - \cos \theta_B = 1, \,\, {\rm CS},
\nonumber \\
&& \theta_A = \theta_B \,\,{\rm and}\,\, \sin \theta_A \neq 0, \,\, {\rm SF},
\nonumber \\
&& \theta_A \neq \theta_B \,\,{\rm and}\,\, \sin \theta_A \neq 0 \,\,{\rm and} \,\,
\sin \theta_B \neq 0, \,\, {\rm CSS},
\nonumber \\
&&\cos \theta_A = \cos \theta_B = \pm 1, \,\,{\rm MI}.
\end{eqnarray}
Minimizing the mean-field energy,
\begin{eqnarray}
\frac{\mathcal{H}_0}{M} &=&
-J\sin\theta_A \sin\theta_B
+\frac{V}{4}(C_{+} + C_{-})\cos\theta_A \cos\theta_B
\nonumber \\
&& 
+\frac{V}{8}(C_{+} - C_{-})(\cos^2 \theta_A + \cos^2\theta_B)
\nonumber \\
&&
- \frac{h}{4}(\cos\theta_A + \cos\theta_B),
\end{eqnarray}
with respect to $\theta_A$ and $\theta_B$, we obtain the phase diagram in the 
$(V/J,h/J)$-plane as shown in Fig.~\ref{fig:PD}(a).
Reflecting the particle-hole symmetry of the hardcore Bose-Hubbard model, the phase
diagram is symmetric with respect to the line $h=0$.

In Fig.~\ref{fig:PD}(a), it is seen that  the CSS and CS phases are present when $V/J>2/C_{-}$.
Increasing $h/J$ from the $n=0$ MI region with a fixed value of $V/J > 2/C_{-}$,
the system exhibits the continuous transitions to SF at $h=h_{\rm SF}^{c}$, CSS at 
$h=h_{\rm CSS}^{c}$, and CS at $h=h_{\rm CS}^{c}$ in order.
This behavior is clearly illustrated in Fig.~\ref{fig:PD}(b), where the filling factor $n$ is 
plotted as a function of $h/J$.
The critical values of $h$ for these transitions are given by
\begin{eqnarray}
h_{\rm SF}^c &=& \pm (4J +2C_{+}V),
\nonumber \\
%
h_{\rm CSS}^c &=& \pm 2(C_{+}V+2J)\sqrt{\frac{C_{-}V-2J}{C_{-}V+2J}},
\nonumber \\
%
h_{\rm CS}^c &=& \pm 2 \sqrt{(C_{-}V)^2 - 4J^2}.
\label{eq:critH}
\end{eqnarray}
We note that replacing $C_{\pm}V$ with $2(V_1\pm V_2)$, Eq.~(\ref{eq:critH}) coincides 
with the critical values of $h$ obtained in previous work for the hardcore Bose-Hubbard
model with the nearest-neighbor interaction $V_1$ and the next-nearest-neighbor
interaction $V_2$~\cite{scalettar-95,pich-98}.
When $V/J$ is increased further, there emerge different solid and SS phases 
in addition to the phases described above.
For instance, allowing for the four-sublattice density modulation, we calculate the 
boundary to the SS2 phase as shown in Fig.~\ref{fig:PD}(a).
The SS2 phase is sketched in Fig.~\ref{fig:PD}(II).
We do not push our calculations  into the region of $V>5.2J$, where other solid 
and SS phases are present, because our purpose is to investigate
superfluidity of the SF and CSS phases.
Notice that there also exist numerous meta-stable states in the region of large 
$V/J$~\cite{menotti-07}, which make experimental investigation of the ground-state 
phase diagram practically very difficult.

\begin{figure}[tb]
\includegraphics[scale=0.45]{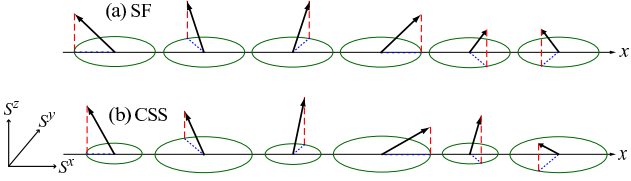}
\caption{\label{fig:flow}
(color online)
Schematic pictures of the SF and CSS states with superflow in the spin representation.
The arrows represent the local direction of the spins.
The radius of the circles denotes the projection of the spins onto the $xy$-plane, which 
corresponds to the local condensate density $n_j^{\rm con}$.
The red dashed lines denote the projection onto the $z$-axis corresponding to the local
density $n_j$. 
}
\end{figure}
\begin{figure}[tb]
\includegraphics[scale=0.45]{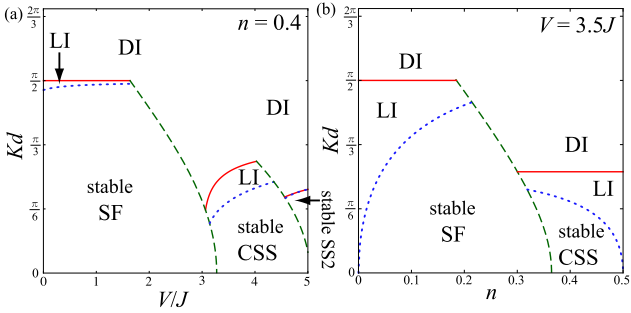}
\caption{\label{fig:SPD}
(color online)
Stability phase diagrams of hardcore bosons flowing in a 2D lattice with 
quasi-momentum ${\bf K}=(K,0)$.
In (a) the regions of stable SF, stable CSS, stable SS2, Landau instability (LI), and 
dynamical instability (DI) are located when $V/J$ is varied for $n=0.4$,
while in (b) those are located when the filling factor $n$ is varied for $V=3.5J$.
The dotted blue, solid red, and dashed green lines represent the critical quasi-momenta
for LI, DI caused by phonons, and DI caused by roton-like excitations.}
\end{figure}

\vspace{3mm}
\section{Excitation spectra and critical velocity}
\label{sec:Vc}
Having established the location of the SF and CSS phases in the phase diagram,
we next study stability of superflow in these phases by means of a linear 
stability analysis.
Let us consider that the optical lattice confining hardcore bosons is moving at a constant
velocity ${\bf v}$.
In the coordinate system where the lattice is at rest, the SF component of the hardcore
bosons is flowing with quasi-momentum ${\bf K}=-m{\bf v}$~\cite{sarlo-05,mun-07},
where $m$ is the particle mass.

In the SF phase, a current-carrying solution of Eqs.~(\ref{eq:static1}) and (\ref{eq:static2}) 
is given by $\bar{\theta}_j = \theta_0$ and $\bar{\varphi}_j = - {\bf K}\cdot {\bf r}_j$, where
$\theta_0$ is related to the filling factor as $\cos \theta_0 = 2n-1$.
This state is sketched in Fig.~\ref{fig:flow}(a).
The current carried by this state is given by 
${\bf j}=2n(1-n)J\sum_{m=x,y}{\bf e}_m \, \sin(K_m d)$, where $(K_x,K_y)\equiv {\bf K}$, and ${\bf e}_x$ and ${\bf e}_y$
represent the unit vectors in the $x$ and $y$ directions. 
Inserting this solution into Eqs.~(\ref{eq:MFenergy}) and (\ref{eq:static2}), 
we obtain the energy per particle
$\epsilon_{\bf K} \equiv (\mathcal{H}_0 + h \sum_j \langle S_j^z \rangle)/N$ and 
the chemical potential $h_{\bf K}$ as functions of ${\bf K}$:
\begin{eqnarray}
\epsilon_{\bf K} = -4(1-n)J\gamma_{\bf K} +\frac{(1-2n)^2}{2n}C_{+}V,
\label{eq:Esf}
\end{eqnarray}
\begin{eqnarray}
h_{\bf K} = 2(2n-1)(2J\gamma_{\bf K} + C_{+}V).
\end{eqnarray}
where $\gamma_{\bf K} = \sum_{m=x,y}\cos (K_m d)/2$.
Notice that $\epsilon_{\bf K}$ and $h_{\bf K}$ satisfy the thermodynamic relation, 
$h_{\bf K} = \frac{\partial (N\epsilon_{\bf K})}{\partial N}$.

Solving Eqs.~(\ref{eq:bogo1}) and (\ref{eq:bogo2}), we obtain the excitation spectrum,
\begin{widetext}
\begin{eqnarray}
\omega_{\bf K}({\bf q}) \! = \! (2 \! - \! 4n)J(\gamma_{{\bf q}-{\bf K}} \! - \! \gamma_{{\bf q}+{\bf K}})
\! + \! \left[
4J(2\gamma_{\bf K} \! - \! \gamma_{{\bf q} + {\bf K}} \! - \! \gamma_{{\bf q} - {\bf K}})
\!
\left\{
2J\gamma_{\bf K} \! - \! (1 \! - \! 2n)^2
J(\gamma_{{\bf q} + {\bf K}} \! + \! \gamma_{{\bf q}-{\bf K}})
\! + \! n(1\! - \! n)V({\bf q})
\right\}
\right]^{\frac{1}{2}}
\!\! ,
\label{eq:exSF}
\end{eqnarray}
\end{widetext}
where ${\bf q}$ is the quasi-momentum of the excitation and
%
$
V({\bf q})/V = 
\sum_{\alpha=1}^{\infty}\frac{2}{\alpha^3}
\left\{ \cos(\alpha q_x d) + \cos(\alpha q_y d )
\right\}
+ \sum_{\alpha_x, \alpha_y =1}^{\infty} 
\frac{4}{(\alpha_x^2 + \alpha_y^2)^{3/2}}
\cos(\alpha_x q_x d)\cos(\alpha_y q_y d).
$
%
Stability of superflow in the SF state can be judged by Eq.~(\ref{eq:exSF}).
We assume that the current is flowing in the $x$-direction, i.e. ${\bf K}=(K,0)$, and
depict the stability phase diagrams in the $(n,Kd)$- and $(V/J, Kd)$-planes in
Fig.~\ref{fig:SPD}.
Stability of the hardcore boson system deep in the SF region, e.g. 
$V\lesssim J$ or $0< n \ll 0.5$, is analogous to that of softcore boson systems described 
by the GP mean-field theory~\cite{smerzi-02, menotti-03} in the following way.
With increasing $K$, excitations with $\omega_{\bf K}({\bf q})<0$ appear at a certain value 
of the quasi-momentum, $K=K^{\rm L}_{\rm SF}$, signaling LI.
In Fig.~\ref{fig:SPD}, $K^{\rm L}_{\rm SF}$ is plotted by the dotted blue line separating 
the stable SF and LI regions.
When $K>K^{\rm D}_{\rm SF} = \pi/(2d)$, long-wavelength phonons cause DI.
This DI reflects the fact that the effective mass in the $x$-direction, 
defined by $(\tilde{m}_{\bf K}^{x})^{-1}=\frac{\partial^2 \epsilon_{\bf K}}{\partial K_x^2}$,
is negative, resulting in the imaginary sound speed in the $x$ direction
$c_{\bf K}^x = (\kappa_{\bf K} \tilde{m}_{\bf K}^{x})^{-1/2}$.
Here, $\kappa_{\bf K}$ is the compressibility.

In the SF region close to the boundary with the CSS phase, the excitation spectrum
$\omega_{\bf K}({\bf q})$ has a roton-like minimum at 
${\bf q}={\bf k}_{\pi}$~\cite{scalettar-95}.
When $K$ increases in this region, the roton-like excitations cause DI, which signals the 
transition to the CSS phase~\cite{zhao-06,burkov-08,ganesh-09,yunomae-09}, before LI occurs.
The critical value of $K$ for this DI is plotted by the dashed-dotted green line 
in Fig.~\ref{fig:SPD}.
Previous theoretical work for the system of hardcore bosons with only 
nearest-neighbor interaction has predicted that superflow can be destabilized 
by the roton-like excitations, but that the resulting CSS state with superflow is 
dynamically unstable~\cite{burkov-08,ganesh-09}.
In contrast, we will show below that the flowing CSS state can be stable in our system
of dipolar hardcore bosons.

\begin{figure}[t]
\includegraphics[scale=0.45]{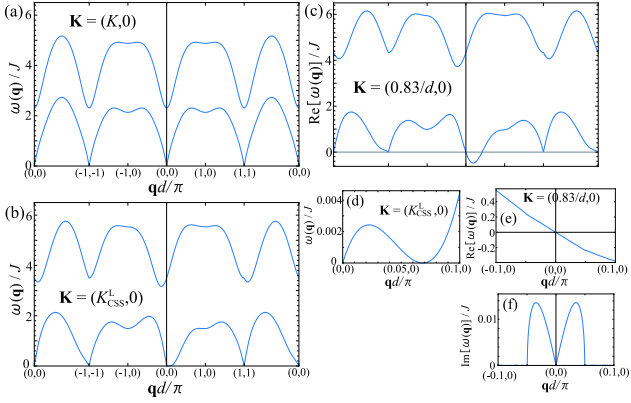}
\caption{\label{fig:excitation}
(color online)
Excitation spectra $\omega({\bf q})$ in the CSS phase for $V=3.5J$, $n=0.4$, 
and different values of $K$ in the $x$-direction, where $K=0$ (a), 
$K=K^{\rm L}_{\rm CSS}$ (b), and $K=0.83/d>K^{\rm D}_{\rm CSS}$ (c).
(d) and (e) are magnification of (b) and (c) focused on the region of
$|q_x|\ll 1/d$ and $q_y=0$, where excitations with negative and 
complex energies arise. 
In (f), the imaginary part of $\omega({\bf q})$ corresponding to (e) is shown.
}
\end{figure}

In the CSS phase, solving Eqs.~(\ref{eq:static1}) and (\ref{eq:static2}) within the 
two-sublattice ansatz, we obtain a current-carrying solution,
\begin{eqnarray}
\!\!\!\!\!\!\!\!\!\!\!\!\!
&&\bar{\varphi}_j = -{\bf K}\cdot {\bf r}_j, \, \cos\theta_A = 4n-2-\cos\theta_B,
\nonumber \\
\!\!\!\!\!\!\!\!\!\!\!\!\!
&&\cos\theta_B = 2n-1\pm 2 \sqrt{n^2-n+\frac{1}{2}-\frac{|1-2n|C_{-}V}{2\beta_{\bf K}}},
\label{eq:CSSflow}
\end{eqnarray}
where $\beta_{\bf K} = \sqrt{(C_{-}V)^2-(2J\gamma_{\bf K})^2}$.
A schematic picture of this state is depicted in Fig.~\ref{fig:flow}(b).
This state possesses the current
${\bf j} = 2|1-2n|J^2\gamma_{\bf K} \beta_{\bf K}^{-1} \sum_{m=x,y}{\bf e}_m \, \sin(K_m d)$.
Substituting Eq.~(\ref{eq:CSSflow}) into Eqs.~(\ref{eq:MFenergy}) and (\ref{eq:static2}), 
we obtain the energy per particle and the chemical potential,
\begin{eqnarray}
\!\!\!\! \epsilon_{\bf K} \!\!&=&\!\! \frac{|1-2n|}{n}\beta_{\bf K} 
+ \frac{(1-2n)^2(C_{+}-C_{-})-C_{-}}{2n} V,
\label{eq:Ecss}
\end{eqnarray}
\begin{eqnarray}
h_{\bf K} = {\rm sgn}(2n-1)\times 2\beta_{\bf K} - 2(1-2n)(C_{+} - C_{-})V.
\end{eqnarray}

In Fig.~\ref{fig:excitation}, assuming ${\bf K}=(K,0)$ again, we show the excitation spectra
for $V=3.5J$, $n=0.4$, and the different values of $K$.
There are two branches of excitation spectrum: one is a gapless
and linear mode at low momenta, denoted by $\omega_{\bf K}^{-}({\bf q})$, and the other is 
a gapful mode, denoted by $\omega_{\bf K}^{+}(\bf q)$.
Since $\omega^{\pm}_{\bf K} ({\bf q}) \geq 0$ until $K$ exceeds a certain critical value,
in the stability phase diagrams of Fig.~\ref{fig:SPD} there is a region where the CSS
state with finite superflow is stable.
Like in the SF phase, we find two scenarios regarding the instability of superflow
in the CSS phase.
First, sufficiently away from the boundary to the SS2 phase, $\omega_{\bf K}^{-}({\bf q})$ is 
pushed down with increasing $K$, and it reaches zero at $K=K^{\rm L}_{\rm CSS}$ 
(see Figs.~\ref{fig:excitation}(b) and (d)), signaling LI.
In Fig.~\ref{fig:SPD}, $K^{\rm L}_{\rm CSS}$ is plotted by the dotted blue lines 
separating the LI and stable CSS regions.
When $K$ is increased further, DI caused by phonons at low momenta sets in 
at $K=K^{\rm D}_{\rm CSS}$ as seen in Figs.~\ref{fig:excitation}(c), (e), and (f).
$K^{\rm D}_{\rm CSS}$ can be determined by the condition $\tilde{m}_{\bf K}^x=0$, 
which is reduced to
\begin{eqnarray}
(C_{-}V)^2(2\cos(Kd)-1)=J^2(1+\cos (Kd))^2\cos (Kd).
\label{eq:dicss}
\end{eqnarray}
When $C_{-}V\gg J$, we obtain
\begin{eqnarray}
K^{\rm D}_{\rm CSS} d \simeq \frac{\pi}{3}-\frac{3\sqrt{3}}{8}\frac{J^2}{(C_{-}V)^2}.
\end{eqnarray}
In Fig.~\ref{fig:SPD}, the right solid red lines represent $K^{\rm D}_{\rm CSS}$.
It is obvious from Eq.~(\ref{eq:dicss}) and Fig.~\ref{fig:SPD} that $K^{\rm D}_{\rm CSS}$ 
is independent on $n$ and monotonically increases with $V/J$.
It is also worth stressing that $K^{\rm D}_{\rm CSS}$ is distinctively smaller than
$K^{\rm D}_{\rm SF}$.
We attribute this reduction of the critical quasi-momenta to the difference between 
the energy band structures in the SF and CSS phases, which are respectively given by 
Eqs.~(\ref{eq:Esf}) and (\ref{eq:Ecss}).

Secondly, near the boundary to the SS2 phase, DI caused by the excitations at
${\bf q}=(\pi/d,0)$ precedes the other instabilities and signals the transition to the SS2 phase.
The critical value of $K$ for this DI is plotted by the dashed green line in Fig.~\ref{fig:SPD}(a).
To complete the stability phase diagram, we also carry out 
stability analyses for the SS2 state with superflow and locate the 
stable, LI, and DI regions.
Notice that the LI region is almost invisible in Fig.~\ref{fig:SPD}(a) because the critical 
quasi-momentum for LI is nearly equal to that for DI.
%
%

Finally, we discuss the feasibility of measuring experimentally the critical quasi-momenta.
Experiments of ultracold dipolar Bose gases confined in a moving optical lattice will 
be performed in an additional parabolic trap.
Since the density vanishes at the edge of the trapped gas, the critical quasi-momentum
for LI is zero.
Hence, for observing superflow, the temperature has to be so low that the thermal
component is invisible and LI cannot destabilize the 
system~\cite{sarlo-05,konabe-06,iigaya-06}.
Another consequence of the parabolic trap is that the CSS phase inevitably coexists with
other phases if it is present in the trap.
For instance, when $n_j=0.4$ at the trap center and $V=3.5J$, the CSS phase occupies
the central region of the trap and is surrounded by the SF phase.
We recall that the critical quasi-momentum $K_{\rm CSS}^{\rm D}$ for the DI in CSS
is smaller than that in SF and independent on the density.
This means that even in the trapped CSS phase coexisting with the SF phase, one can
observe dissipationless flow by moving the optical lattice with a velocity smaller than
$K_{\rm CSS}^{\rm D}/m$ and the breakdown of the superflow when 
$K_{\rm CSS}^{\rm D}/m$ is exceeded.
Thus, given that the central density and $V/J$ are precisely controllable in experiments,
the stability phase diagrams of Fig.~\ref{fig:SPD} can be investigated in current experimental
setups.

\section{Summary}
\label{sec:Sum}
In conclusion, we have studied stability of superflow of dipolar Bose gases
in a moving optical lattice.
Specifically focusing on the superfluid (SF) and checkerboard supersolid (CSS) phases,
we have calculated the critical quasi-momenta for Landau and dynamical instabilities.
Superflow in the CSS phases has been found to be stable until the quasi-momentum
exceeds the critical value, which is significantly smaller than that in the SF phase.
In the CSS phase, we also found the dynamical instability caused by roton-like excitations that
results in the transition to another type of supersolid.
We emphasize that measuring the critical quasi-momenta will be a direct signature of superfluidity
of the SS phases.

Let us make brief comments on two recent experiments~\cite{vengalattore-10,baumann-10} 
that have explored supersolid phases in different contexts from dipolar bosons in optical lattices.
One of them has studied spin textures in spin-1 Bose-Einstein condensates (BEC) 
of $^{87}$Rb atoms with the ferromagnetic contact interaction and the dipole-dipole 
interaction~\cite{vengalattore-10}.
It was found that the spinor condensate has spatial magnetic order.
However, this experiment is not convincing enough to proclaim the discovery of a supersolid 
phase not only because the magnetic order is short-ranged, but also because the equilibration
time of the system is so long that one can not judge whether the magnetically ordered state
is really an equilibrium state.
The other experiment of Ref.~\cite{baumann-10} has studied the system of a BEC coupled
with an optical cavity and observed the formation of checkerboard density wave order in the 
BEC associated with the superradiant phase transition.
Since the superfluidity of this possible supersolid phase has not been confirmed yet,
it will be important to investigate the critical velocity in this system both theoretically and
experimentally.

{\it Note added:} After the submission of the present paper, there appeared relevant work 
by Kunimi {\it et al.}, which studies the critical velocity of a supersolid phase through a single 
barrier potential~\cite{kunimi-10}.

%
\begin{acknowledgments}
We thank Y. Kato for useful comments. 
The authors are supported by a Grant-in-Aid from JSPS.
\end{acknowledgments}


\begin{thebibliography}{99}
\bibitem{footnote}
Recent theoretical work has proposed a way to measure the superfluid fraction in cold 
atom experiments with the use of a synthetic vector potential for atoms~\cite{cooper-10}, 
but no experiment has tried it yet.
\end{thebibliography}
\end{document}